\documentclass[english,aps,notitlepage,nofootinbib,tightenlines,11pt]{revtex4-1}
\pdfoutput=1
\usepackage[T1]{fontenc}
\usepackage[latin9]{inputenc}
\usepackage{color}
\usepackage{amsmath}
\usepackage{amssymb}
\usepackage{stmaryrd}
\usepackage{graphicx}
\usepackage{esint}

\makeatletter
\usepackage[percent]{overpic}
\usepackage{slashed}
\usepackage{hyperref}
\hypersetup{
    colorlinks,
    linkcolor={blue},
    citecolor={blue},
    urlcolor={blue}
}

\makeatother

\usepackage{babel}
\begin{document}

\title{Wolfenstein potentials for neutrinos induced by ultra-light mediators}

\author{Alexei Yu. Smirnov${}^{a,b}$ and Xun-Jie Xu${}^{a}$}

\affiliation{\textcolor{black}{${}^{a}$ Max-Planck-Institut f\"ur Kernphysik, Postfach 103980, D-69029 Heidelberg, Germany.~\\
${}^{b}$ International Centre for Theoretical Physics, I-34100 Trieste, Italy.}}

\date{\today}
\begin{abstract}
New physics can emerge at low energy scales, involving very light
and very weakly interacting new particles.
These particles can mediate interactions
between neutrinos and usual matter and contribute to the Wolfenstein potential  relevant for neutrino oscillations.
We compute the Wolfenstein potential in the presence of ultra-light scalar and 
vector mediators and study 
the dependence of the potential
on the mediator mass  $m_A$, taking the finite size of matter distribution  (Earth, Sun, supernovae) into consideration.
For ultra-light mediators with $m_{A}^{-1}$ comparable to the size of the medium ($R$), the usual $m_{A}^{-2}$ dependence of 
the potential is modified. In particular, when $m_{A}^{-1}\gg R$,
the potential does not depend on
$m_{A}$. Taking into account existing bounds
on light mediators, we find that for the scalar case
significant  effects on neutrino propagation are not possible, while for the
vector case  large matter effects are allowed
for $m_{A} \in [2\times10^{-17}$, $4\times10^{-14}]$ 
eV and the gauge coupling $g\sim 10^{-25}$.
\end{abstract}
\maketitle

\section{Introduction}
Coherent forward scattering of neutrinos on particles of medium $\psi$ 
($\psi=e^{-}$, $n$, $p$) generates the  Wolfenstein 
potential $V_W$~\cite{Wolfenstein:1977ue}. Being added 
to the neutrino evolution equation, $V_W$ can significantly affect 
neutrino oscillations, known as the Mikheyev-Smirnov-Wolfenstein (MSW) effect~\cite{Wolfenstein:1977ue,Mikheev:1986gs,Mikheev:1986wj}. When  neutrino-matter interactions are mediated by a heavy boson with the  interaction radius $m_{A}^{-1}$ (where $m_{A}$ is the mediator mass) much smaller than 
the object in which neutrinos propagate (or the distance over which the density 
varies), the Wolfenstein potential equals 
\begin{equation}
V_{W}=\frac{g_{\nu}g_{\psi}}{m_{A}^{2}}n_{\psi},
\label{eq:poteninf}
\end{equation} 
where $n_{\psi}$ is the number density of the $\psi$ particles, $g_{\nu}$ and $g_{\psi}$ are the couplings
of  mediator to $\nu$ and $\psi$ respectively. 
The potential depends on the local number density $n_{\psi}$ while the size and shape of the object are not relevant. The medium 
can be considered as infinite. 
In the standard model (SM), the mediators are the $W$ and $Z$ bosons,  
satisfying the condition for (\ref{eq:poteninf}). 
New heavy particles beyond the SM can generate via
non-standard interactions additional contributions to the Wolfenstein potential
with the same form as (\ref{eq:poteninf}).

New neutrino interactions may be mediated by light particles as well,
if the light mediators	are very weakly coupled to the SM fermions. With sufficiently small values of $g_{\nu}$ and $g_{\psi}$, and correspondingly small $m_A$,  sizable  $g_{\nu}g_{\psi}/m_A^2$  can evade  various bounds from processes with large momentum transfer $|q^2|\gg m_A^2$, because the new physics contributions in such processes are typically proportional to the small quantity $g_{\nu}g_{\psi}/|q^2|$ rather than $g_{\nu}g_{\psi}/m_A^2$. 
In contrast, the Wolfenstein potential in (\ref{eq:poteninf}), can be unchanged if $m_A^2$ decreases proportionally with respect to
$g_{\nu}g_{\psi}$. 
This, however, is restricted by the finite 
size of the object $R$.  
When $m_A$ becomes smaller than $1/R$,
the dependence of $V_W$ on $m_A$ in (\ref{eq:poteninf}) 
is modified so that the matter effect turns out to be also  
suppressed. In this paper we will consider this dependence and 
its implications in details. 

The matter effects due to light mediators have 
been studied before 
\cite{Joshipura:2003jh,Grifols:2003gy,Bandyopadhyay:2006uh,
GonzalezGarcia:2006vp,
Nelson:2007yq,GonzalezGarcia:2008wk,Samanta:2010zh,Heeck:2010pg,
Davoudiasl:2011sz,Lee:2011uh,Chatterjee:2015gta,
Bustamante:2018mzu,Khatun:2018lzs,Wise:2018rnb,
Ge:2018uhz,Krnjaic:2017zlz,Berlin:2016woy,Brdar:2017kbt}.
The mediators, mostly considered in the literature, are new gauge bosons of the lepton numbers $L_{e}-L_{\mu}$, $L_{\mu}-L_{\tau}$ or $L_{\tau}-L_{e}$. Long-range forces induced by these bosons 
can affect solar and atmospheric neutrino oscillations  
\cite{Joshipura:2003jh,Grifols:2003gy} as well as 
high energy astrophysical neutrinos interacting
with electrons in the Universe \cite{Bustamante:2018mzu}.
Various fifth force and gravitational experiments put very strong bounds 
on couplings of light mediators with matter but in certain ranges it is   neutrino oscillation phenomena 
that have the best sensitivity to couplings \cite{Wise:2018rnb}.
As for scalar interactions, it is well known that the corresponding matter effect leads to corrections to the neutrino masses.
Recently this possibility has been studied in Ref.~\cite{Ge:2018uhz} 
and it is claimed that such scalar interactions can
explain the discrepancy between the solar neutrino
and KamLAND determinations of $\Delta m_{21}^2$.
%

In this paper, we present detailed study  of 
the Wolfenstein potentials induced by
light mediators (both vector and scalar). 
We compute the Wolfenstein potentials for  several spherically symmetric density profiles and study dependence of the potentials on the mediator mass. 
Taking into account existing bounds on light mediators, we 
assess their relevance to neutrino experiments.
 
The paper is organized as follows. In Sec.~\ref{sec:potential} we
study the effects of light scalar and
vector mediators on neutrino propagation, considering general matter density distributions. 
In Sec.~\ref{sec:specific} we present derivation of the effective potentials 
for several spherically symmetric density distributions which can be applied to the Earth, the Sun and similar celestial bodies. 
In Sec.~\ref{sec:Phenomenology}
we consider existing bounds on  light mediators  and
apply our results to neutrinos propagating in the Sun, the Earth and supernovae.
Discussion and conclusions are presented in Sec.~\ref{sec:Conclusion}.

\section{Effects of light mediators on neutrino propagation \label{sec:potential}}

Let us consider  interactions between neutrinos ($\nu$) and particles in
matter ($\psi$) mediated by a new light vector
boson $A^{\mu}$ or scalar boson $\phi$. The relevant part of the Lagrangian reads  
\begin{equation}
    {\cal L}\supset\overline{\nu}i\slashed{\partial}\nu- 
    m_{\nu}\overline{\nu}\nu-g_{\nu}\overline{\nu}\slashed{A}\nu-
    g_{\psi}\overline{\psi}\slashed{A}\psi+\frac{m_{A}^{2}}{2}A^{\mu}A_{\mu}
    \label{eq:w-A}
    \end{equation}
in the vector case. In the case of a scalar mediator, 
the last three  terms in (\ref{eq:w-A}) should
be replaced by
\vspace{-5mm}
\begin{equation}
{\cal L}\supset-y_{\nu}\overline{\nu}\phi\nu- 
y_{\psi}\overline{\psi}\phi\psi-\frac{m_{\phi}^{2}}{2}\phi^{2}.\label{eq:w}
\end{equation}
We assume that neutrinos are Dirac particles. For Majorana neutrinos,
though the interaction  forms are slightly different, the results
are the same. Also we consider neutrinos of a single flavor. 
It can be  straightforwardly generalized to the case of three-neutrino mixing.

The Lagrangian  (\ref{eq:w-A}) determines the equations of motion (EOM) of  $\nu$ and $A^{\mu}$ 
(in the Lorenz gauge $\partial_{\mu}A^{\mu}=0$):
\begin{eqnarray}
 &  & i\slashed{\partial}\nu-m_{\nu}\nu-g_{\nu}\slashed{A}\nu=0,
\label{eq:w-1-1}\\
 &  & \left[\partial^{2}+m_{A}^{2}\right]A^{\mu}- 
g_{\nu}\overline{\nu}\gamma^{\mu}\nu- 
g_{\psi}\overline{\psi}\gamma^{\mu}\psi=0.
\label{eq:w-3}
\end{eqnarray}

According to Eq.~(\ref{eq:w-1-1}), the effect of $A^{\mu}$ on  neutrino propagation can be described as the displacement
$i\slashed{\partial}\rightarrow i\slashed{\partial}-g_{\nu}\slashed{A}$,
which in the momentum space corresponds to 
\begin{equation}
p^{\mu}\rightarrow p^{\mu} + g_{\nu}A^{\mu},
\label{eq:w-p}
\end{equation}
where $p^{\mu}$ is  the 4-momentum of the neutrino. 
In particular,  the neutrino energy $E$ receives the correction:
\begin{equation}
E=p^{0}\rightarrow E+V,\ \ \ V = g_{\nu}A^{0}.
\label{eq:w-12}
\end{equation}

In the scalar case, the EOM from Eq.~(\ref{eq:w}) are 
\begin{eqnarray}
 &  & i\slashed{\partial}\nu-m_{\nu}\nu-y_{\nu}\phi\nu=0,\label{eq:w-1}\\
 &  & \left[\partial^{2} + m_{\phi}^{2}\right]\phi - 
y_{\nu}\overline{\nu}\nu - y_{\psi}\overline{\psi}\psi=0.
\label{eq:w-2}
\end{eqnarray}
As follows from Eq.~(\ref{eq:w-1}), the effect of $\phi$ on neutrino propagation
is equivalent to changing the neutrino mass:  
\begin{eqnarray}
m_{\nu} & \rightarrow & m_{\nu}+\delta m_{\nu},\ \ \  
\delta m_{\nu}=y_{\nu}\phi.
\label{eq:w-m}
\end{eqnarray}

In most  applications, the medium particles $\psi$ 
are at rest (non-relativistic), hence\footnote{Recall that the physical meaning of  $\overline{\psi}\gamma^{\mu}\psi$
is the electric current density  and $\overline{\psi}\gamma^{0}\psi=\psi^{\dagger}\psi$
is the electron number density \cite{Peskin}. For $\psi$ at rest, $\overline{\psi}\psi=\overline{\psi}\gamma^{0}\psi$ is identical to the electron number density.}
\begin{equation}
\overline{\psi}\psi=n_{\psi},\ \overline{\psi}\gamma^{\mu}\psi= 
n_{\psi}(1,\ 0,\ 0,\ 0).
\label{eq:w-13}
\end{equation}
Since the neutrino number density is much smaller than the number
density of electrons or nucleons, 
we take $\overline{\nu}\nu\ll\overline{\psi}\psi$ and 
$\overline{\nu}\gamma_{\mu}\nu\ll\overline{\psi}\gamma_{\mu}\psi$ in Eqs.~(\ref{eq:w-2}) and (\ref{eq:w-3}).
This means that $\phi$ and $A^{\mu}$ are dominantly induced by $\psi$.

For the vector case,  Eq.~(\ref{eq:w-13}) implies that the spatial
components of  $A^{\mu}$ vanish  (up to gauge uncertainties):
\begin{equation}
A^{\mu}=(A^{0},\ 0,\ 0,\ 0).\label{eq:w-6}
\end{equation}
Furthermore, since the $\psi$ particles  are at rest, $A^{0}$
has no temporal dependence ($\partial_{t}A^{0}=0$).  Therefore,
Eq.~(\ref{eq:w-3}) becomes
\begin{equation}
\left[-\nabla^{2}+m_{A}^{2}\right]A^{0}=g_{\psi}n_{\psi}.
\label{eq:w-4}
\end{equation}
Given a distribution of $n_{\psi}$, Eq.~(\ref{eq:w-4}) determines  $A^{0}$.

All the above analyses can be straightforwardly applied to a scalar
mediator. Starting from Eq.~(\ref{eq:w-2}), we obtain the equation
similar to Eq.~(\ref{eq:w-4}):
\begin{equation}
\left[-\nabla^{2}+m_{\phi}^{2}\right]\phi=y_{\psi}n_{\psi},
\label{eq:w-27-1}
\end{equation}
and hence a similar solution.

From Eqs.~(\ref{eq:w-4})  and (\ref{eq:w-27-1}), we can see that $\phi$ and $A^{0}$ are determined by $n_{\psi}$ in the same way. 
However, the effects of $\phi$ and $A^{0}$ on neutrino propagation
are very different.

In the neutrino evolution equation
$\phi$ gives correction to the mass and therefore appears as
$(m_\nu + \delta m_\nu)^2/2E$.
Approximately(for $\delta m_{\nu}\ll m_{\nu}$), this corresponds to adding
\begin{equation}
V_S \approx  \frac{m_\nu}{E}\delta m_\nu  = \frac{m_\nu}{E} y_\nu \phi
\label{eq:scalar}
\end{equation}
to $(m_\nu)^2/2E$. 
In comparison, $A^0$ appears in the equation as an
addition of $V = g_\nu A^0$ to $m_\nu^2/2E$.
Thus, the scalar matter effect enters the flavor evolution equation with 
the additional suppression factor $m_{\nu}/E$. 
This factor is due to chiral suppression. 
Since the helicity is conserved in neutrino oscillations,  the
chirality-flipping terms in the Lagrangian, such as  mass terms or scalar interactions, have to change it twice, which is the reason that $m_{\nu}^2$ appears instead of $m_{\nu}$  in neutrino oscillations.
Because the scalar interaction flips the chirality, another flip is needed which is given by $m_{\nu}/2E$. In other words, only $m_{\nu}/2E$ fraction of the 
chirality-flipped state contains the original helicity. 
This means that, 
to obtain effects of the same size,  
$y_\nu \phi$ should be $\frac{E}{m_\nu}$ times larger than  $g_\nu A^0$. That is,  $y_\nu y_\psi /m_\phi^2$ should be  $\frac{E}{m_\nu}$
times larger than $g_\nu g_\psi/m_A^2$. This strongly affects the relevance of the scalar case to the oscillation phenomenology.

In the case of pseudo-scalar and axial-vector mediators,
the corresponding fields are produced by interactions
with $\overline{\psi}\gamma^5\psi$ and $\overline{\psi}\gamma^{\mu}\gamma^5\psi$.  In unpolarized medium, these two quantities
vanish due to $\gamma^5$. For polarized medium,
the pseudo-scalar and axial-vector potentials decrease as $r^{-3}$,
being dipole effects (see, e.g.,  Ref.~\cite{Chikashige:1980ui}). Therefore these potentials are small compared to scalar 
and vector potentials.

\section{Effective potentials for spherically symmetric density distributions\label{sec:specific}}
\vspace{0.3cm}
In what follows, we consider the vector case, while  for the scalar case, the results can be obtained immediately with the substitutions: $A^{0}\rightarrow\phi$, $m_{A}\rightarrow m_{\phi}$,
$g_{\nu}\rightarrow y_{\nu}$, and $g_{\psi}\rightarrow y_{\psi}$. 

In many applications (e.g., the Earth, the Sun), the matter density distribution is, to a good approximation,  spherically symmetric.  The   spherical symmetry allows us to reduce Eq.~(\ref{eq:w-4}) to a radial differential equation:
\begin{equation}
\left[\frac{\partial^{2}}{\partial r^{2}}+\frac{2\partial}{r\partial r}-
m_{A}^{2}\right]A^{0}(r)=-g_{\psi}n_{\psi}(r).\label{eq:w-5}
\end{equation}
For any given profile $n_{\psi}(r)$, Eq.~(\ref{eq:w-5}) can  be solved by 
a standard method known as \emph{variation of parameters}, 
which gives 
\begin{equation}
A^{0}(r)=\frac{g_{\psi}}{m_{A}r}\left[e^{-m_{A}r}\int_{0}^{r}xn_{\psi}(x)\sinh(m_{A}x)dx+\sinh(m_{A}r)\int_{r}^{\infty}xn_{\psi}(x)e^{-m_{A}x}dx\right].\label{eq:A_general}
\end{equation}

The above computations of potentials are essentially 
classical. Therefore  for  an arbitrary  
density distribution $n_\psi(\boldsymbol{r})$ (not necessarily spherically 
symmetric),   $A^0$ can be found by performing summation (integration) of
the Yukawa potentials produced by individual particles
[see also Eq.~(\ref{eq:w-yuakwa})]: 
\begin{equation}
A^{0}(\boldsymbol{r})=
-\frac{g_{\psi}}{4\pi}\int n_{\psi}(\tilde{\boldsymbol{r}})\frac{e^{-m_{A}|
\boldsymbol{r} - 
\tilde{\boldsymbol{r}}|}}{|\boldsymbol{r} - 
\tilde{\boldsymbol{r}}|}d^{3}\tilde{\boldsymbol{r}}.
\label{eq:A_int}
\end{equation}
For the spherically symmetric case, one can integrate over angular variables in Eq.~(\ref{eq:A_int}), which also leads to Eq.~(\ref{eq:A_general}). 

Using Eq.~(\ref{eq:A_general}) we compute the Wolfenstein potentials for several
matter density profiles, 
which can be used to approximately describe the density distributions of the Earth, the Sun, and supernovae. Some interesting limits will be discussed.


\vspace{0.5cm}
{\bf \noindent $\blacksquare$ \ Constant density within a sphere}
\vspace{0.5cm}

For a constant density distribution within a sphere  of radius $R$:  
\begin{equation}
n_{\psi}(r)=\begin{cases}
0 & ({\rm for\ }r>R)\\
n_{\psi} & ({\rm for\ }r\leq R)
\end{cases},
\label{eq:w-15}
\end{equation}
we  find from (\ref{eq:A_general})
\begin{equation}
A^{0}(r)=\frac{g_{\psi}n_{\psi}}{m_{A}^{2}}F(r),\label{eq:w-7}
\end{equation}
where
\begin{equation}
F(r)=\begin{cases}
1-\frac{m_{A}R+1}{m_{A}r}e^{-m_{A}R}\sinh(m_{A}r) & \ (r\leq R)\\[2mm]
\frac{e^{-m_{A}r}}{m_{A}r}\left[m_{A}R\cosh(m_{A}R)
-\sinh(m_{A}R)\right] & \ (r>R)
\end{cases},
\label{eq:w-F}
\end{equation}
describes deviation from the infinite medium potential.
Then according to Eq.~(\ref{eq:w-12}), the effective neutrino potential produced
by $A^{0}$ equals 
\begin{equation}
V(r)=g_{\nu}A^{0}(r)=\frac{g_{\nu}g_{\psi}n_{\psi}}{m_{A}^{2}}F(r).
\label{eq:w-V}
\end{equation}

\vspace{0.4cm}

\noindent Several important limits are in order.
\vspace{5pt}

\noindent$\bullet$ $R\rightarrow0$.
We fix the number of particles $\psi$, $N_{\psi}\equiv\frac{4}{3}\pi R^{3}n_{\psi}$, within the sphere
when taking
$R\rightarrow0$. In this limit $n_\psi(r) \rightarrow N_\psi \delta(r)$, Eqs.~(\ref{eq:w-F}) and (\ref{eq:w-V}) give
\begin{equation}
V(r)=g_{\nu}g_{\psi}
\frac{e^{-m_{A}r}}{r}n_{\psi}\frac{R^{3}}{3}
=g_{\nu}g_{\psi}N_{\psi}\frac{e^{-m_{A}r}}{4\pi r},
\label{eq:w-yuakwa}
\end{equation}
which, for $N_{\psi}=1$, reproduces the Yukawa potential  of a single $\psi$ particle.

\vspace{5pt}

\noindent$\bullet$ $m_{A}\rightarrow0$. 
In the limit of massless mediator, 
Eqs.~(\ref{eq:w-F}) and (\ref{eq:w-V}) lead to 
\begin{equation}
V(r)=g_{\nu}g_{\psi}n_{\psi}\times\begin{cases}
\frac{3R^{2}-r^{2}}{6} & \ (r\leq R)\\[2mm]
\frac{R^{3}}{3r} & \ (r>R)
\end{cases}\label{eq:w-7-1}
\end{equation}
which coincides with the $r$ dependence of the Coulomb potential
of a charged sphere. 
In this limit, the potential for $r > R$ is determined by the total particle number $N_{\psi}$ inside the sphere, independently of the radial distribution. 
If $m_{A}$ is nonzero but small, the first
order correction in $m_{A}$ to the potential equals (for both $r\leq R$
and $r>R$)
\begin{equation}
\delta V(r)=-g_{\nu}g_{\psi}n_{\psi}\frac{m_{A}}{3}R^{3}.
\label{eq:w-18}
\end{equation}

\vspace{5pt}

\noindent$\bullet$ $m_{A}\rightarrow\infty$. 
In this limit, Eqs.~(\ref{eq:w-F}) and (\ref{eq:w-V}) give
\begin{equation}
V(r)=\begin{cases}
\frac{g_{\nu}g_{\psi}n_{\psi}}{m_{A}^{2}} & \ (r\leq R)\\[2mm]
0 & \ (r>R)
\end{cases}.\label{eq:w-7-1-1}
\end{equation}
Inside medium, the potential is a constant and outside it vanishes.
This reproduces the standard Wolfenstein potential
for infinite medium.

\vspace{5pt}

\noindent$\bullet$ $|R-r|\ll m_{A}^{-1}\ll R$.
The limit means that the radius of interaction is much smaller
than $R$ but much larger than the depth of the trajectory $R-r$.
The absolute sign ``$|\ |$'' indicates that this limit applies
not only to underground neutrino trajectories but also to above-the-surface
neutrino beams. The limit can be realized for reactor and
accelerator experiments.  
In this limit,  we obtain 
\begin{equation}
V(r)=\frac{1}{2}\frac{g_{\nu}g_{\psi}n_{\psi}}{m_{A}^{2}}.\label{eq:w-19}
\end{equation}
It differs from the standard Wolfenstein potential by the  additional factor
of $1/2$, which reflects that only half of the space produces the potential.

\begin{figure*}[t]
\centering
\includegraphics[width=0.45\textwidth]{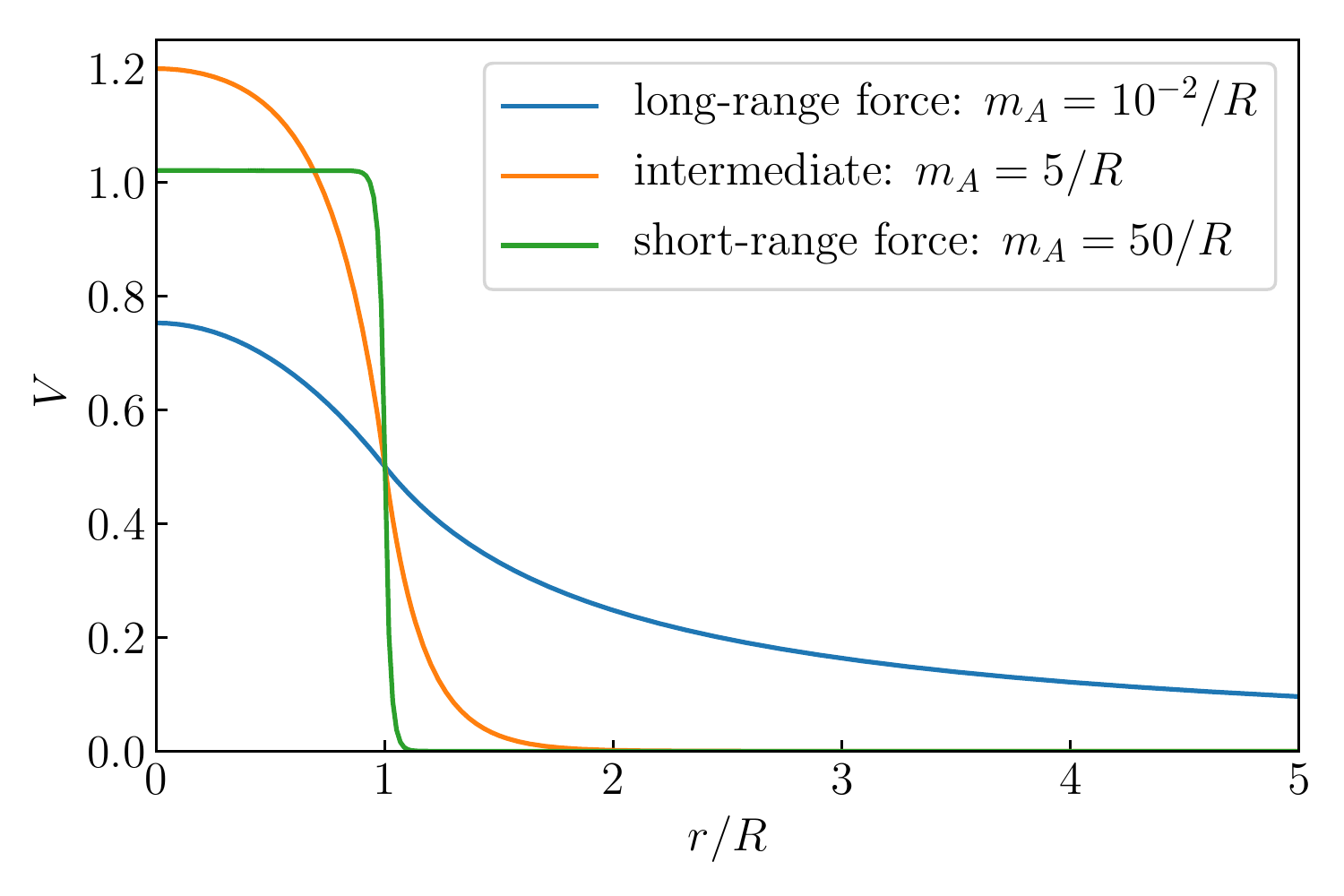}\ \ 
\includegraphics[width=0.45\textwidth]{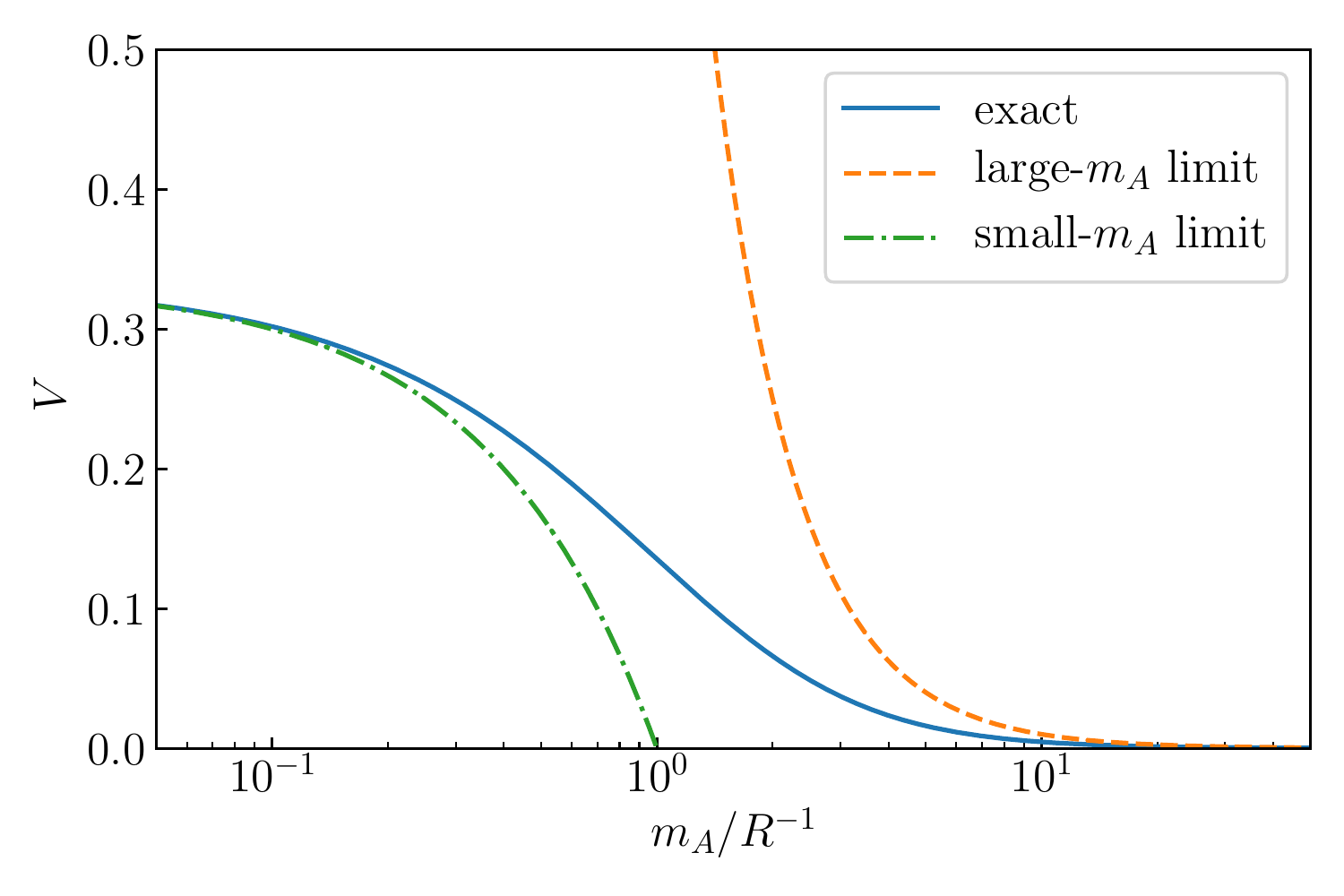}
\caption{\label{fig:V_curves}
The effective potentials $V$ (in arbitrary unit) produced 
by a sphere of radius $R$ with constant matter density. {\it Left panel:}
Dependence of $V$ on the distance $r$ for different values of the
mediator mass $m_A$. Values of $g_{\nu}g_{\psi}n_{\psi}$
are chosen in such a way that $V(r=R)=1/2$ for all values of $m_{A}$. 
{\it Right panel:} Comparison 
of $V(m_A)$  in the large and small $m_A$
limits  with the exact result, computed using Eqs.~(\ref{eq:w-7})-(\ref{eq:w-V}) and (\ref{eq:w-7-1})-(\ref{eq:w-7-1-1}) at the
surface of the sphere ($r=R$). 
}
\end{figure*}

\vspace{0.5cm}
In the left panel of Fig.~\ref{fig:V_curves}, we show the dependence
of $V$ on $r$ for different values of $m_{A}$ 
according to (\ref{eq:w-V}) and (\ref{eq:w-F}). For
$m_{A}\gg R^{-1}$ (green curve) the $r$-dependence is close to
that of the standard Wolfenstein potential, which is essentially a
step function. 
Near the surface ($r\approx R$) the short-range
potential (green curve) is roughly half the standard Wolfenstein potential
(the plateau of this curve), as expected from Eq.~(\ref{eq:w-19}). 
When $m_{A}$ decreases, it becomes smoother
(yellow line). For very small $m_{A}$, the long-range force leads to a Column-like
potential (blue curve).  

In the right panel of Fig.~\ref{fig:V_curves}, we show the dependence
of $V$ at $r=R$  on  $m_{A}$  according to the
exact formula (blue solid curve) as well as the small- and large-$m_{A}$
limits  in Eqs.~(\ref{eq:w-7-1})-(\ref{eq:w-7-1-1}). 
The inverse-square dependence of the standard Wolfenstein
potential on $m_A$ deviates significantly
from the exact dependence for small values of $m_A$.

\vspace{0.5cm}
{\bf \noindent $ \blacksquare $  Multi-layer density profile }
\vspace{0.5cm}

A multi-layer density profile with constant densities
within layers is a good approximation
for the Earth density distribution.
Using result  (\ref{eq:w-F}) for constant density spheres one can obtain
results for the Earth.
For instance, for two layers (a simplified mantle-core profile) 
with constant average densities $n_\psi^M$ and  $n_\psi^C$ and  radii
$R^M$ and $R^C$ respectively,   the Wolfenstein potential in the mantle is
\begin{equation}
V(r)=\frac{g_{\nu}g_{\psi}}{m_{A}^{2}}
\left[n_\psi^M F_{r < R}(r, R^M) + (n_\psi^C - n_\psi^M) F_{r > R}(r, R^C)\right],
\label{eq:earth-mantle}
\end{equation}
and in the core is
\begin{equation}
V(r)=\frac{g_{\nu}g_{\psi}}{m_{A}^{2}}
\left[n_\psi^M F_{r < R}(r, R^M) + (n_\psi^C - n_\psi^M) F_{r < R}(r, R^C)\right],
\label{eq:earth-core}
\end{equation}
where $F_{r < R}$ and  $F_{r > R}$ are given by the lower and the upper 
lines of Eq.~(\ref{eq:w-F}).  For more realistic profiles with many layers, the generalization is straightforward.

\vspace{0.5cm}
{\bf \noindent $\blacksquare$ \ Exponential density distribution}
\vspace{0.5cm}

Exact analytic results can be obtained for an exponential density distribution, 
\begin{equation}
n_{\psi}(r)=n_{\psi}(0)e^{-r\kappa}.
\label{eq:w-15-1}
\end{equation}
This distribution can be used to partially describe the matter density of the Sun and supernovae.
Computing the integral in (\ref{eq:A_general}) gives 
\begin{equation}
V(r)=\frac{g_{\nu}g_{\psi}n_{\psi}(r)}{m_{A}^{2}- 
\kappa^{2}}\left[1+\frac{2\kappa}{m_{A}^{2} - 
\kappa^{2}}\frac{1}{r}\left(e^{(\kappa-m_A)r}-1\right)\right].
\label{eq:w-21}
\end{equation}

\noindent 
Several important limits are in order. 

\vspace{5pt}

\noindent$\bullet$ $\kappa\rightarrow0$.
This limit corresponds to a constant density distribution. 
In this limit, Eq.~(\ref{eq:w-21}) reduces to 
the standard Wolfenstein potential of infinite medium, see Eq.~(\ref{eq:w-7-1-1}) with $ n_\psi = n_\psi(0)$.

\vspace{5pt}

\noindent$\bullet$ $m_{A}\ll\kappa$. 
In this limit, 
 the radius of  force is much larger
than the scale of density change.
From  Eq.~(\ref{eq:w-21}), we obtain
\begin{equation}
V(r)=-\frac{g_{\nu}g_{\psi}n_{\psi}(r)}{\kappa^{2}}
\left[1-\frac{2}{\kappa r}\left(e^{\kappa r}-1\right)\right],
\label{eq:w-21-1}
\end{equation}
which  does not depend on $m_A$. 
For $r\gg\kappa^{-1}$, Eq.~(\ref{eq:w-21-1}) reduces to 
\begin{equation}
V(r)=\frac{g_{\nu}g_{\psi}n_{\psi}(0)}{\kappa^{2}}\frac{2}{\kappa r},
\label{eq:w-24}
\end{equation}
that is, the potential at
 large $r$ decreases as $1/r$, 
i.e., slower than the matter density ($\propto e^{-\kappa r}$) decreases.
This can produce interesting phenomena such as new level crossing for solar neutrinos \cite{as-xu}.

\vspace{5pt}

\noindent$\bullet$ $m_{A} \approx \kappa$. 
In this limit, Eq.~(\ref{eq:w-21})
gives
\begin{equation}
V(r)=\frac{g_{\nu}g_{\psi}n_{\psi}(r)}{4\kappa^{2}}(1+\kappa r).\label{eq:w-26}
\end{equation}

\vspace{0.5cm}
{\bf \noindent $\blacksquare$ \ Exponential density distribution with a cut-off}
\vspace{0.5cm}

The density profile of the Sun can be
described more accurately than in (\ref{eq:w-15-1})
by an exponential  distribution with a cut-off at the solar radius:
\begin{equation}
n_{\psi}(r)=n_{\psi}(0)\begin{cases}
e^{-r\kappa} & (r\leq R)\\
0 & (r>R)
\end{cases}.\label{eq:wn-28}
\end{equation}
After straightforward calculations, we obtain the potential:
\begin{equation}
V(r)=\frac{g_{\nu}g_{\psi}n_{\psi}(0)}{m_{A}r}\times\begin{cases}
K_{{\rm in}} & (r\leq R)\\
K_{{\rm out}} & (r>R)
\end{cases},\label{eq:wn-25}
\end{equation}
with $K_{{\rm in}}$ and $K_{{\rm out}}$ given by 
\begin{eqnarray}
K_{{\rm in}} & \equiv & \kappa m_{A}e^{-r(\kappa+m_{A})}\frac{e^{m_{A}r}\left(m_{A}^{2}r\kappa^{-1}-\kappa r-2\right)+2e^{\kappa r}}{(m_{A}^{2}-\kappa^{2})^{2}}\nonumber \\
 & - & \frac{\sinh(m_{A}r)e^{-R(\kappa+m_{A})}(m_{A}R+\kappa R+1)}{(m_{A}+\kappa)^{2}},\label{eq:wn}
\end{eqnarray}
\begin{eqnarray}
K_{{\rm out}} & \equiv & e^{-m_{A}r-\kappa R}\sinh(m_{A}R)\frac{m_{A}^{2}(\kappa R-1)-\kappa^{2}(\kappa R+1)}{\left(m_{A}^{2}-\kappa^{2}\right)^{2}}\nonumber \\
 & + & e^{-m_{A}r-\kappa R}\cosh(m_{A}R)\frac{m_{A}^{3}R-\kappa^{2}Rm_{A}-2\kappa m_{A}}{\left(m_{A}^{2}-\kappa^{2}\right)^{2}}+e^{-m_{A}r}\frac{2\kappa m_{A}}{\left(m_{A}^{2}-\kappa^{2}\right)^{2}}.\label{eq:wn-10}
\end{eqnarray}

All the above results can be applied to the scalar case with the simple substitution: $A^{0}\rightarrow\phi$, $m_{A}\rightarrow m_{\phi}$,
$g_{\nu}\rightarrow y_{\nu}$, and $g_{\psi}\rightarrow y_{\psi}$. 

\section{Phenomenology\label{sec:Phenomenology}}
 
\begin{figure*}[!tp]
\centering
\includegraphics[width=0.48\textwidth]{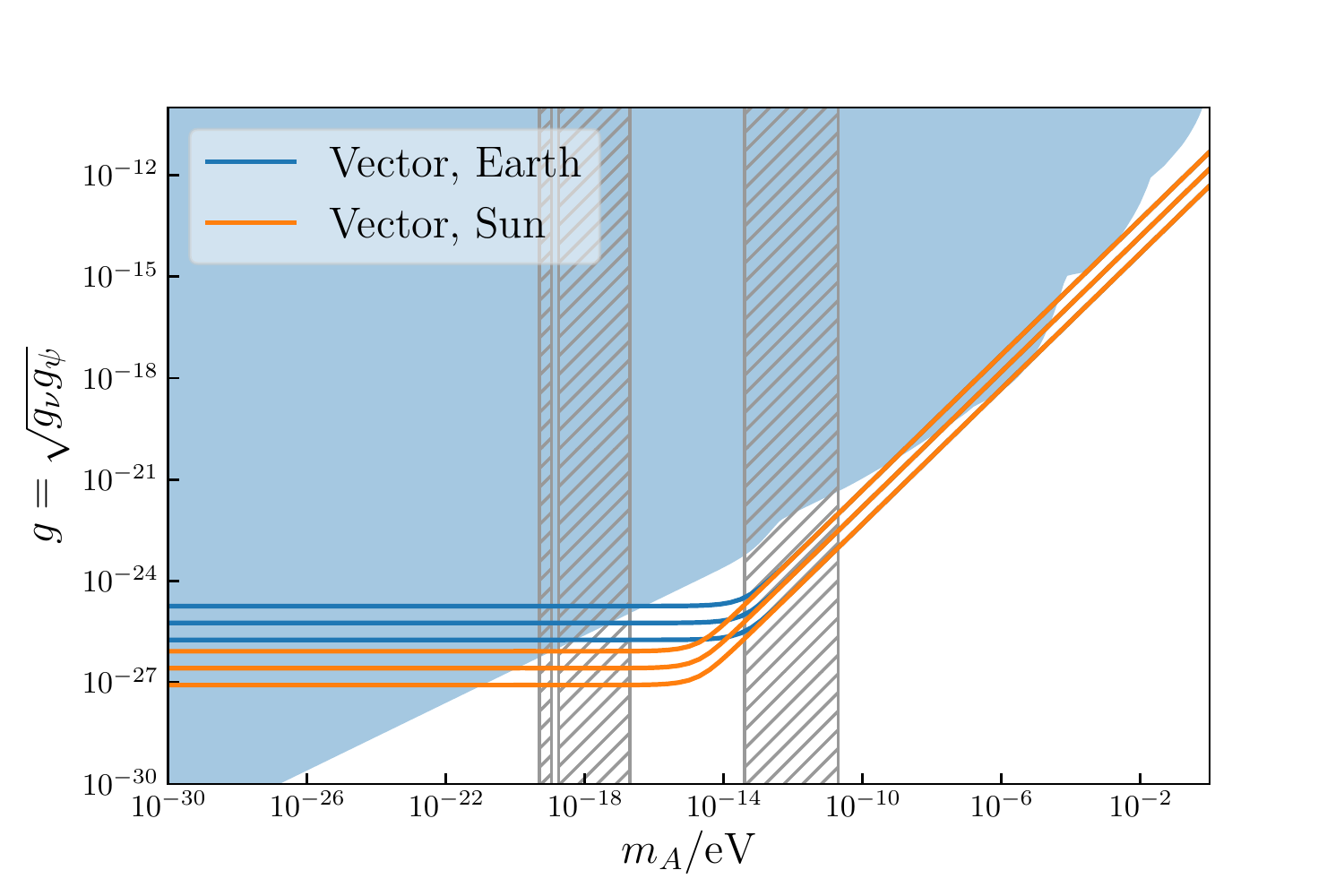}\ \ 
\includegraphics[width=0.48\textwidth]{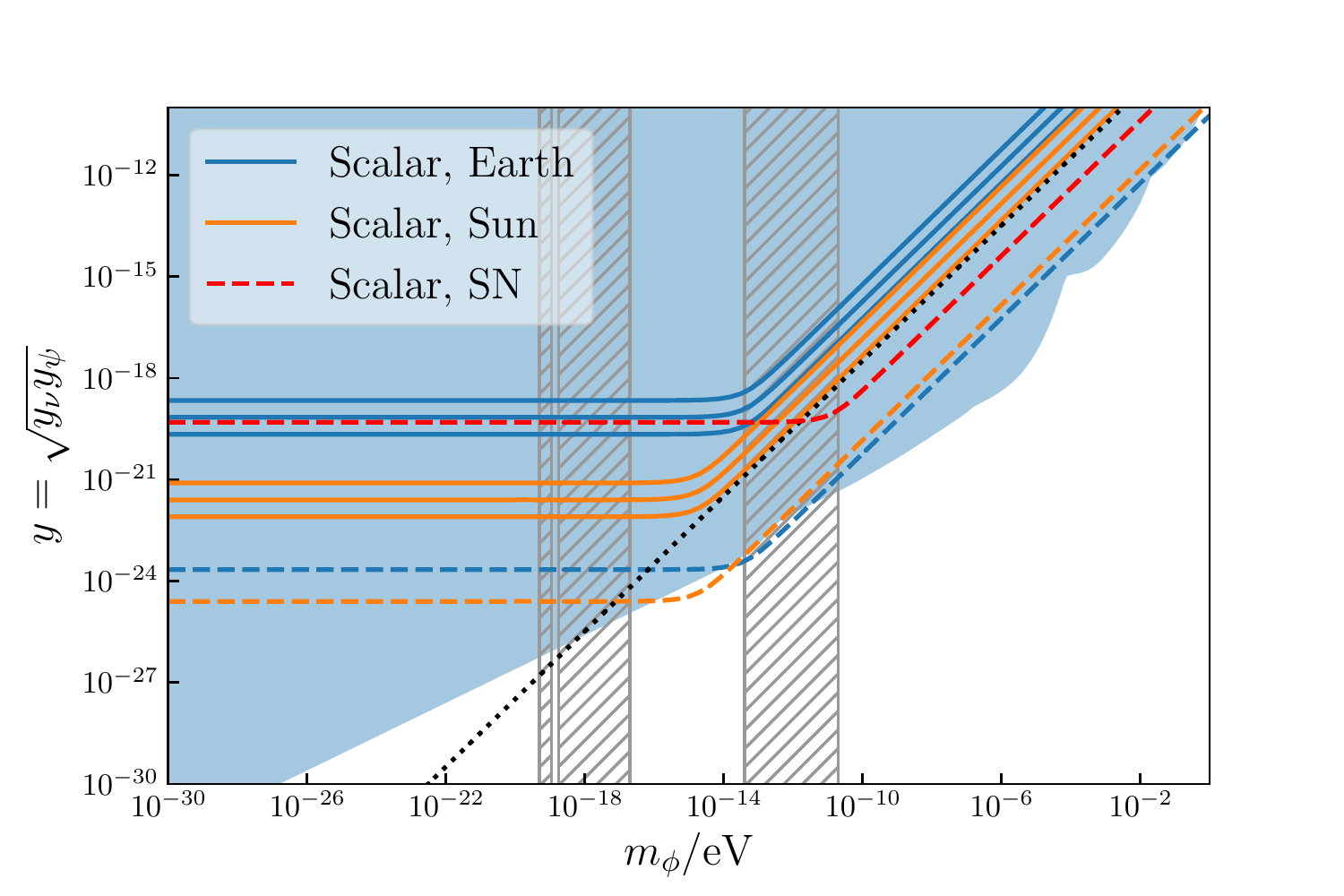}
\caption{
The regions of masses and couplings of ultra-light vector
(left) and scalar (right) mediators in which significant
effects are produced by the matter of the Earth at its surface
(blue lines) and by matter of the Sun in its center (orange lines).
The blue shaded regions are excluded by the combination
of various observations (explained in the text).
The grey hatched regions are excluded by black hole super-radiance.
Left panel: Lines of equal potentials generated by vector mediator.
From down to up: 
$V/V_{{\rm SM}}=\{10^{-2},\ 10^{-1},\  1\}$.
Right panel: Lines of equal corrections to the neutrino mass
generated by scalar mediator. 
From down to up:
$\delta m_{\nu}/{\rm eV}=\{10^{-3},\ 10^{-2},\ 10^{-1}\}$.
For comparison, we plot two dashed lines
with $\delta m_{\nu}=10^{-11}$ eV for the Earth and 
$\delta m_{\nu}=10^{-8}$ eV for the Sun
which have parts below the excluded regions.
Red dashed line corresponds to  $\delta m_{\nu} = 5$  MeV
generated by the core of supernovae. The region above
this line is excluded (see text).
\label{fig:bounds}
}
\end{figure*}

Let us consider experimental bounds on the couplings and masses of light
mediators and check whether light scalar and vector mediators, that
satisfy these bounds, can produce observable matter effects in neutrino
oscillations. For brevity we introduce the couplings  
$g \equiv \sqrt{g_{\nu}g_{\psi}}$ and $y \equiv \sqrt{y_{\nu}y_{\psi}}$.

For $m_{A,\phi}\gtrsim{\cal O}(10^{2})$ keV, 
$g^2/m_A^2\gtrsim G_{F}$ and $ y^2/m_\phi^2 \gtrsim G_{F}$
have been excluded by numerous laboratory experiments  including the elastic
neutrino-electron scattering \cite{Rodejohann:2017vup,Lindner:2018kjo,Arcadi:2019uif}, 
neutrino-nucleus 
scattering \cite{Lindner:2016wff,Farzan:2018gtr,Brdar:2018qqj}, fixed target experiments \cite{Bjorken:2009mm,Batell:2009di,Essig:2010gu},
collider searches \cite{Lees:2014xha,TheBABAR:2016rlg}, etc.  For
$1\ {\rm eV}\lesssim m_{A,\phi}\lesssim10^{2}\ {\rm keV}$, the astrophysical
observations provide much stronger constraints. 
For instance, the expected amount of energy
loss via neutrinos in the Sun and globular clusters 
excludes 
$g$ and $y$
down to $10^{-14}$, corresponding to  an upper bound of
$g^2/m_A^2$ or $\ y^2/m_\phi^2$ above $10^{-5}G_{F}$ \cite{Harnik:2012ni}.

Below 1 eV, the constraints mainly come from searches of fifth
forces and precision tests of gravity \cite{Adelberger:2006dh,Schlamminger:2007ht}.
These constraints are only applicable to $g_{\psi}$ or
$y_{\psi}$. To obtain the bounds on $\sqrt{g_{\nu}g_{\psi}}$ or
$\sqrt{y_{\nu}y_{\psi}}$, we can use the cosmological bounds on neutrino
self-interactions, $g_{\nu}^{2}/m_{A}^{2}$ or $y_{\nu}^{2}/m_{\phi}^{2}$
$\lesssim(3\ {\rm MeV})^{-2}$
\cite{Kreisch:2019yzn}. In addition, for certain ranges of $m_{A,\phi}$,
the black hole super-radiance provides robust constraints \cite{Baryakhtar:2017ngi,Bustamante:2018mzu}
which are independent of the couplings.  These  constraints
are combined and presented in Fig.~\ref{fig:bounds}. 

Notice  that these constrains for the mass of  mediators  
below 1 eV (shown in Fig.~\ref{fig:bounds})  are based on tests of gravity 
and cosmological observations, and therefore flavor independent. 
Searches for the fifth force and precision tests of gravity are
sensitive to the couplings of mediators  with matter $g_{\psi}$ ($y_{\psi}$), 
while the bounds on neutrino couplings $g_{\nu}$ ($y_{\nu}$)  
are based on the effect of neutrino free streaming on CMB,   
which are independent of neutrino flavors. 
Above 1 eV, some of the mentioned constraints,  e.g., from neutrino scattering, do depend on  
flavors. However,  studies in this region of masses 
show that new interactions cannot be much larger than the SM interactions.
Therefore above 1 eV the scalar case remains excluded,
whereas the vector case still has viable parameter space, as has been
widely discussed in the literature.  

Now let us determine the required values of  
$g$ or $y$
to generate significant matter effects. The latter can be quantified by
\begin{equation}
\frac{V}{V_{\rm SM}} = \frac{g^2}{m_A^2 \sqrt{2} G_F} F(r, m_A^2),\ \ 
\delta m_{\nu} = \frac{y^2}{m_\phi^2} n_{\psi}(r) F(r, m_\phi^2),
\label{eq:w-23}
\end{equation}
where $V_{{\rm SM}}\equiv  \sqrt{2} G_{F}n_{\psi}$, 
and $F$ presented in Eqs.(\ref{eq:w-F}) describes 
the deviation from the standard Wolfenstein potential.  
According to Eq.~(\ref{eq:w-23}), 
for given values of $\frac{V}{V_{\rm SM}}$ and $\delta m_{\nu}$, $g^2$ and $y^2$ are determined by:
\begin{equation}
g^2 = \frac{V}{V_{\rm SM}} m_A^2 \sqrt{2} G_F F^{-1}(r, m_A^2),\ \ \ 
y^2 = \delta m_{\nu} {m_\phi^2} [ n_{\psi(r)} F(r, m_\phi^2)]^{-1} .
\label{eq:w-zz}
\end{equation}
In Fig.~\ref{fig:bounds}, we plot the dependence $g = g(m_A)$ and
$y = y(m_{\phi})$ from Eq.~(\ref{eq:w-zz}) for  
$V/{V_{\rm SM}} \in \{10^{-2},\ 10^{-1},\ 1\}$ and 
$\delta m_{\nu} \in \{10^{-3},\ 10^{-2},\ 10^{-1}\}$ eV
correspondingly.  The curves determine the bands of strong
(observable) matter effects in the mediator parameter space.
The values of $F$ have been computed numerically
for the center of the Sun ($r = 0$) using the solar density
distributions from \cite{Edsjo:2017kjk},
and for the surface of the Earth ($r=R_{\oplus}$) using the density distribution
from \cite{Dziewonski:1981xy}.
The value of $F$ has been computed numerically using 
the density distributions for the Sun from \cite{Edsjo:2017kjk}
and for the Earth from \cite{Dziewonski:1981xy}. The value of $r$ is set to zero for the Sun, and to the Earth radius for the Earth.

The dependence $g = g(m_A)$ and
$y = y(m_{\phi})$
can be understood 
from our analytic results. For large  $m_A$ (short range forces), 
$F \approx 1$. According to Eq.~(\ref{eq:w-zz}), we have $g \propto m_A^2$ for $m_A \ll 1/R$, where $R$ is the radius of the object. 
For very small $m_A$, $F$ should be proportional to $m_A^2$ 
so that $g$ is independent of $m_A$. In the intermediate 
range, $m_A \sim 1/R$,  the dependence is more complicated.

According to Fig.~\ref{fig:bounds}, the solid curves turn at the values
of mediator masses $2.8\times10^{-16}$ eV for the Sun,
and $3.1\times 10^{-14}$ eV for the Earth that are determined
by inverse of the solar radius $R_{\varodot}=7.0\times10^{5}$ km
and the Earth radius $R_{\oplus}\approx6.4\times10^{3}$ km correspondingly.
Below these masses at $m_A, \ m_{\phi} < 10^{-16 }$ eV (for the Sun)
and $m_A, \ m_{\phi} < 10^{-14}$ eV (for the Earth) the curve
becomes horizontal, i.e. dependence on the mediator masses
disappears. This corresponds essentially to the Coulomb-like
potentials generated by massless mediators.


According to Fig.~\ref{fig:bounds}  (left) there are several $m_A$-$g$
regions of observable matter effects which are generated by the vector
mediator and allowed by the present bounds. In the long-range forces
domain they include
\begin{equation}
m_A = (10^{-21} - 10^{-19})\ {\rm eV}, \  \  \  g = (10^{-27} - 
10^{-26}),
\end{equation}
for neutrinos in the Sun, and
\begin{equation}
m_A = (10^{-21} - 10^{-19})\ {\rm eV}, \ \ \   g = (10^{-27} - 
10^{-26}),
\end{equation}
both for neutrinos in the Sun and in the Earth.
The unexcluded regions in the short range forces domain are
the same for the neutrinos in the Sun and the Earth:
\begin{equation}
m_A = (10^{-11} - 10^{-6})\ {\rm eV}, \  \  \  g = (10^{-24} - 
10^{-19}),
\end{equation}
\begin{equation}
m_A > 10^{-4}\  {\rm eV}, \  \  \  g > 10^{-17}.
\end{equation}
This could motivate further phenomenological studies. 

As follows from Fig.~\ref{fig:bounds} (right panel) the scalar mediators
cannot generate significant matter effects,
in contrast to the  claim in Ref.~\cite{Ge:2018uhz}. 
For comparison, a black dotted line corresponds
to $\delta m_\nu = 0.001$ eV generated
in the infinite size medium with the same density
as in the center of the Sun.
(For large $m_\phi$ it coincides with the corresponding
line for the Sun.)
In this case,  significant matter
effects would be possible for $m_{\phi}\lesssim10^{-20}\ {\rm eV}$). 
However, this possibility is excluded  
when $F(r)$ (the  correction) due to the finite size of medium is taken  into account.  

One may be interested in the allowed values of scalar
matter effect. 
So, we plot two dashed curves that
correspond to values of corrections
$\delta m_{\nu}=10^{-11}$ eV for the Earth and $\delta m_{\nu}=10^{-8}$ eV
which have small sections below the excluded region.
These values are far beyond the precision
of any realistic experiments.

Finally, strong constraints on the scalar mediator parameters
can be obtained from supernova neutrinos,
which have not been considered in the literature. Let us estimate the effect  in the core of a supernova with a typical radius of $R_{\rm SN} = 20\sim30$ km and a 
density of $n^{\rm SN}_\psi = 10^{13}\sim10^{14}$
g$/$cm$^{3}$.  For relatively heavy scalar mediators with
$m_\phi> 1/R_{\rm SN}$, the mass correction
\begin{equation}
\delta m_\nu^{\rm SN} \sim
\frac{n^{\rm SN}_\psi}{n^{\rm earth}_\psi}
\delta m_\nu^{\rm earth},
\label{eq:w-SN}
\end{equation}  
is about 13 orders of magnitude larger than 
the correction in the Earth 
$\delta m_\nu^{\rm earth}$. For $\delta m_\nu^{\rm earth} = 0.01$ eV, 
we obtain $\delta m_\nu^{\rm SN} \sim 100$ GeV,
which is certainly excluded since in this case neutrinos
can not even be produced in supernovae.
For very light mediators with $m_\phi < 1/R_{\rm SN}$, Eq.~(\ref{eq:w-SN}) is modified to
\begin{equation}
\delta m_\nu^{\rm SN} \sim
\frac{n^{\rm SN}_\psi}{n^{\rm earth}_\psi}
\left(\frac{R^{\rm SN}}{R^{\rm earth}}\right)^2
\delta m_\nu^{\rm earth}.
\end{equation}
Taking $\delta m_\nu^{\rm earth} = 0.01$ eV, 
we obtain $\delta m_\nu^{\rm SN} = 5$ MeV and correspondingly,
$y = 10^{-20}$ (see Fig.~\ref{fig:bounds}). Since neutrino masses $m_{\nu} > 5$ MeV
would strongly affect production of neutrinos and their energy
spectrum, values of couplings $y > 10^{-20}$ can be excluded.
 More detailed analyses 
 will be given elsewhere \cite{as-xu}.

\section{Conclusion\label{sec:Conclusion}}

We studied the effects of new neutrino-matter interactions mediated by ultra-light scalar and vector bosons on neutrino propagation, 
taking into account the finite size and density distribution of the medium.
We compute the Wolfenstein potentials explicitly for several spherically symmetric density profiles which can be approximately applied to the Earth and the Sun. 

The Wolfenstein potentials induced by ultra-light  mediators have very different dependence on the mediator masses ($m_A$) from the standard case.
In infinite medium,  due to the $1/m_{A}^{2}$ dependence
in Eq.~(\ref{eq:poteninf}), the Wolfenstein potentials can be enhanced by
reducing the mediator masses. 
In finite medium of size $R$, when $m_A$ 
decreases down to the region
where  $m_A^{-1}$ becomes comparable or larger than $R$, the $1/m_{A}^{2}$ dependence  is modified.
In this case the matter effect
is determined by the geometry
(including the size) and density distribution
of the object, rather than the local density.
In particular,  the potential can extend outside of the medium. 
For $m_A\ll R$, the potential does not depend on $m_A$.

With correct expressions for the Wolfenstein potentials and existing bounds on light mediators, we find that
scalar mediators cannot produce 
observable effects in realistic
experiments, which implies that the scenario considered in Ref.~\cite{Ge:2018uhz}
is not viable. Vector mediators, however, 
can produce significant matter effect,  in particular,  within the parameter space $m_A\in${[}$2\times10^{-17}$,
$4\times10^{-14}${]} eV and $g\sim10^{-25}$.


\bibliographystyle{apsrev4-1}
\bibliography{ref}

\end{document}